# ON THE ORBIT OF VISUAL BINARY WDS 01158-6853 I-27CD (SAO248342)

S.Siregar[1,2]

[1]Bosscha Observatory ITB, Lembang 40391
[2]Astronomy Research Division, Bandung Institute of Technology, Indonesia

**Abstract.** WDS 01158-6853 I-27CD=SA0 248342 has the proper motion +404' in right ascension and 105" in declination. Magnitude of each star is 7.84 for primary and 8.44 for secondary, separated by 320' from the quadruple system Kappa Tuc=LDS 42 = HJ 3423 AB. The visual binary star of WDS 01158-6853 I-27CD is historically one of the most important double star in constellation Tucana. We have collected the observational data consisting of separation angular (ρ) and position angle (θ) from the observations of 1897 up to 2001 taken at the Bosscha Observatory and other Observatories in the world. This study presents the recent status of orbit binary system WDS 01158-6853 I-27CD. By using Thiele Van den Bos method and empirical formula of Strand's Mass-Luminosity relation we have determined the orbit and mass of WDS 01158-6853 I-27CD. The results are;

| Dynamical Elements | Orbit Orientation | Masses-Parallax |
|---|---|---|
| P = 85.288 years | i = 27.93 | $M_1 = 0.7\,M_0$ |
| e = 0.053 | Ω = 52.83 | $M_2 = 0.5\,M_0$ |
| T = 1911.23 | ω = 10.73 | p = 0".0589 |

## 1 Introduction

Since the Bosscha Observatory (λ=105⁰ and φ=-6⁰30') was established in 1923 researches on visual binary stars has been playing an important role in astronomical researches in Indonesia for example [10] proposed a method to determine the orbit according to three base points of observation for moderate length of secondary trajectories. The WDS 01158-6853 I-27CD is a visual double star, compose of primary star and secondary star with magnitude respectively mv = 7.84 and mv = 8.44 both of the stars have spectral type of K2V. For knowing the influences of nearest stars in orbital elements we studied this system by using the two body problem and made a comparison to previous studied.

There are many computational techniques proposed by astronomers. Algorithm developed gave us possibility to determine the elements orbit. In this work the Thiele-van den Bos method (vide; [1] ) has been used that is, employing three complete observations of the observing time, angular separation and position angle (t,ρ,θ ) together with the double area constant which must be obtained from additional data. This method already used to determine orbit of visual binary COU 297 [5] other examples; ADS 8119 AB [7], and Alpha Centauri [8]. From the view point of accuracy the observation data considered here were between years 1911 and 2001, collected at Bosscha Observatory and other sources. The figure 1 represents the curve of ρ and θ as function of t

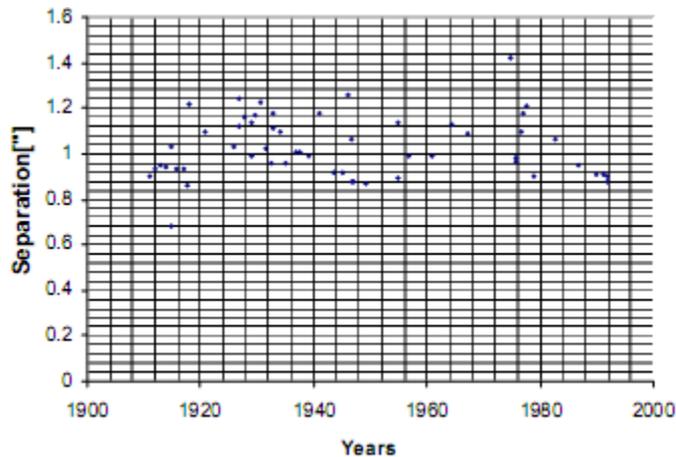

**Figure 1.** Separation angle shown that WDS 01158-6853 I-27CD has completed one revolution

## 2 Data and Notations

The data for this work collected from year 1897 up to 2001 were taken from cataloque of Washington Double Star (WDS). These data presented the star almost one time passed its periastron in average periode around 80 years. To determine the double areal constant C assumed the recent data are more accurate then other, therefore for orbit determination all of the data of (ρ, θ) from epoch 1911 up to 2001, were chosen. These data are shown in figure 1 and figure 2

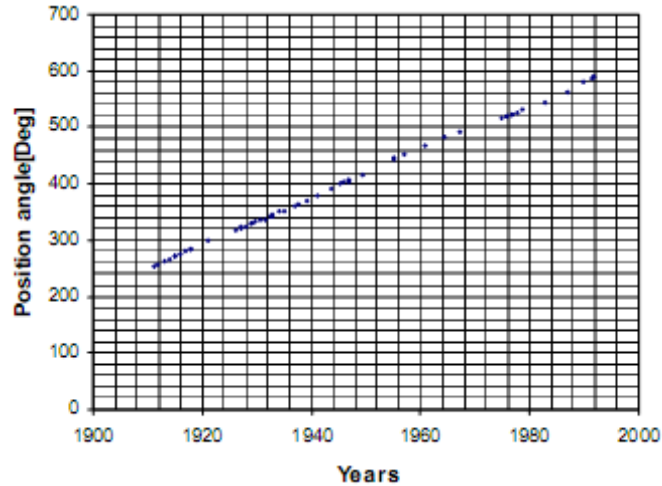

**Figure 2.** Position Angle as function of observing time of WDS 01158-6853 I-27CD from year 1911 up to 2001.The curve shown that the period of system is around 80 years

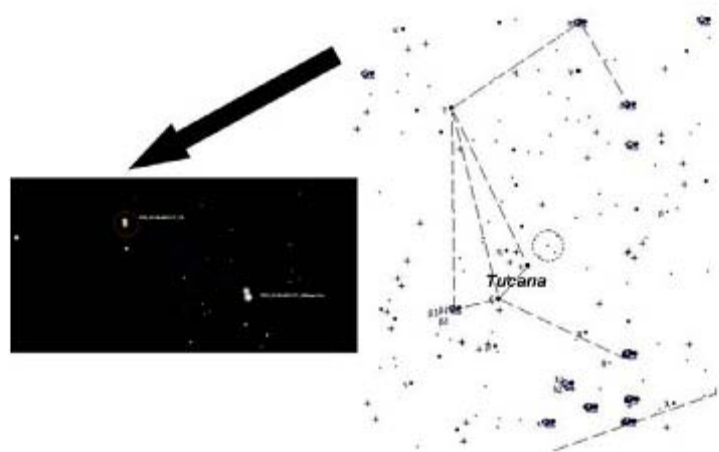

**Figure 3.** The quadruple system of Kappa Tuc=LDS 42 = HJ 3423 AB and WDS 01158-6853 I-27CD=SA0 248342 separated by 320 arc second

In this work we define some symbols;
P,T,e,a,i,ω,Ω have their usual meaning for orbits
μ= 2π/P, is the mean annual motion
$t_i$ = time at which the secondary star occupies its i-th position
$E_i$ = eccentric anomaly at time $t_i$
$t_{ij}$ = $t_j$ - $t_i$
$E_{ij}$ = $E_j$ - $E_i$
V = $E_{12}$ ; U = $E_{23}$ ; W = $E_{13}$
($θ_i$,$ρ_i$) = coordinates of the secondary star at time $t_i$
$Δ_{ij}$ = $ρ_iρ_j$ Sin ($θ_j$-$θ_i$ )
C = the double areal constant of the apparent orbit

We define several supplement variables
$S_1 = (t_{23} - \Delta_{23}/C)$; $S_2 = (t_{13} - \Delta_{13}/C)$; $S_3 = (t_{12} - \Delta_{12}/C)$
$R = \Delta_{12}/\Delta_{13}$; $S = \Delta_{12}/\Delta_{23}$
$M = R(t_{12} - S\, t_{23})/(R\, t_{13} - S\, t_{23})$
$N = S(t12 - R\, t13)/(R\, t_{13} - S\, t_{23})$
$Q = (RS+R-S)^2/2RS$

## 3 Fundamental Equations

For the three pairs (t1,t2), (t2,t3) and (t1,t3) the Thiele-Innes equations ( vide; [1]) can be writen in Kepler's equation. The expressions are;

$$f_1 = U - \sin U - \mu S_1$$
$$f_2 = V - \sin V - \mu S_2$$
$$f_3 = W - \sin W - \mu S_3$$
$$f_4 = U + W - V$$

(3.1)

The three elliptical segments S1,S2 and S3 must be definite positive. In case of an elliptical orbit, the equation (3.1) will lead to a solution for a mean annual motion only when is satisfied;

$$f = S_1^{1/3} - S_2^{1/3} + S_3^{1/3} > 0$$

(3.2)

It should be remembered that function f always increase with increasing C. Hence a slight increase of C may manage a sufficient increase of f, so that an elliptical solution may finally appear.

By choosing mean annual motion µ and regarding the condition U+W=V, the value of U,V and W are determined by iteration method of Newton[4]. To start an iterative process, an initial value µ must be taken around a real mean annual motion. Arend (vide; [6]) has shown that the best choice is to take;

$$\mu = 77.5 \left[ -f(S_1 - S_2 + S_3) \right]^{3/2} e^{-1.9138 f}$$

(3.3)

In case the convergence of the mean annual motion is not reached, i.e. its mean V≠ U+V, the cycle of calculation must be repeated with µ+δµ, U+δU, V+δV and W+δW where δµ, δU, δV and δW are solution of linear systems;

$$\begin{bmatrix} -f_1 \\ -f_2 \\ -f_3 \\ -f_4 \end{bmatrix} = \begin{bmatrix} \frac{\partial f_1}{\partial U} & \frac{\partial f_1}{\partial V} & \frac{\partial f_1}{\partial W} & \frac{\partial f_1}{\partial \mu} \\ \frac{\partial f_2}{\partial U} & \frac{\partial f_2}{\partial V} & \frac{\partial f_2}{\partial W} & \frac{\partial f_2}{\partial \mu} \\ \frac{\partial f_3}{\partial U} & \frac{\partial f_3}{\partial V} & \frac{\partial f_3}{\partial W} & \frac{\partial f_3}{\partial \mu} \\ \frac{\partial f_4}{\partial U} & \frac{\partial f_4}{\partial V} & \frac{\partial f_4}{\partial W} & \frac{\partial f_4}{\partial \mu} \end{bmatrix} \begin{bmatrix} \delta U \\ \delta V \\ \delta W \\ \delta \mu \end{bmatrix}$$

(3.4)

This iterative process is stopped when; $\varepsilon^2 \leq \delta U^2 + \delta V^2 + \delta W^2 + \delta \mu^2$. In this work the tolerance factor ε is less than 10$^{-3}$ radian. Once the relationship between U,V and W has been obtained, the well-known expression;

$$e^2 = 1 + \frac{S(1 - \cos U) + 1 - \cos W - R(1 - \cos V)}{Q}$$

(3.5)

Define e as function of U,V and W. The behaviors of this function and various types of solutions have been published by [2]. For calculating the orbital parameters from U,V,W, µ and e is used the classical method such as;
[U,V,W, µ,e] → [Thiele-Innes elements] → [a,i,ω,Ω]

Through mass-luminosity relation, primary and secondary masses are determined. For detail explanation see [1]. In case of an elliptical motion this mapping is guaranteed, but for non periodic orbits the Thiele-Innes elements must be modified [3]

## 4  Application and Results

As an object study we shall consider of the binary WDS 01158-6853 I-27CD. The orbit for this double star has previously been calculated [9]. In order to make the example as comprehensive as possible, we shall start from three fundamental points and the apparent area constant C = 0.0363 [radian][arcsecond]$^2$ /year. Table.1 gives the three base points and the relative positions of the secondary star of WDS 01158-6853 I-27CD which will be used to determined the mean annual motion

| t | $\theta(°)$ | $\rho('')$ | $t_{qp}$ | $\Delta\theta_{qp}(°)$ | $\Delta_{qp}$ |
|---|---|---|---|---|---|
| 1914 | 266.28 | 0.92 | 38 | 151.56 | 0.4118 |
| 1952 | 417.84 | 0.94 | 38 | 148.56 | 0.4462 |
| 1990 | 566.4 | 0.91 | 76 | 300.2 | 0.7242 |

**Table 1.** Three base points of visual binary WDS 01158-6853 I-27CD

The orbital elements of visual binary WDS 01158-6853 I-27CD computed by this method is presented. A comparison is made with other results in table 2.

| Orbital Elements | This Work | other author[9] |
|---|---|---|
| P[Years] | 85.288 | 85.2 |
| T[Year] | 1911.23 | 1919 |
| µ[deg] | 4.221 | 4.225 |
| a['']  | 1.21 | 1.14 |
| e[.] | 0.053 | 0.04 |
| i[deg] | 27.93 | 35 |
| ω[deg] | 52.83 | 141 |
| Ω[deg] | 10.73 | 142 |

**Table 2.** Orbital elements of WDS 01 158-6853 I-27CD

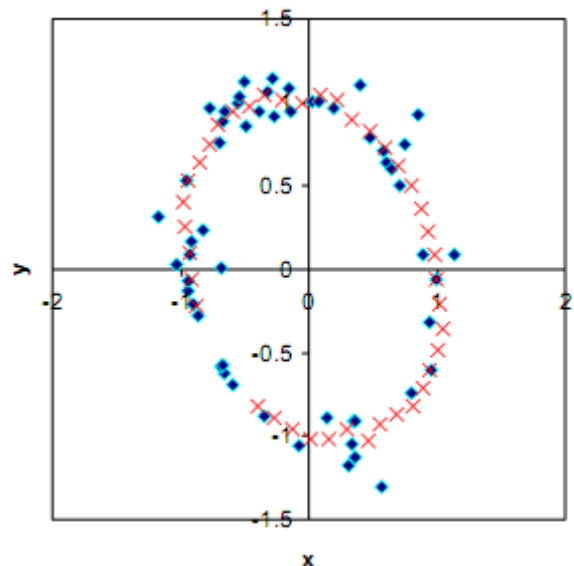

**Figure 4.** The orbit of WDS 01 158-6853 I-27CD on plane x-y. The legend are: x- interpolation (fitting curve) and ◆- observational data

According to the present data in the table the dynamical elements there are not much different, but significant difference founded in i,ω, and Ω. Probably the assumption of Keplerian motion is not sufficient. This preliminary result is needed to be continued by using the n-body problem instead of two-body problem. This system has mass of primary, $M_1$= 0.7 $M_\odot$, and of secondary $M_2$ = 0.5 $M_\odot$, otherwise dynamical parallax p = 0̋.0589 Main while the Thiele -Innes constants follow;

A = -0″.35022,
B = -0″.83141,
F = 1″.03881,
G = -0″.57974

The trajectory of this system is presented in figure 4.

# Acknowledgment

This research has made use of Sixth Catalog of Orbits of Visual Binary Stars data-base operated at US Naval Observatory, Washington, USA (http://ad.usno.mil/wds/orb6.html). The author is indebted to Ms.Nurhasanah for some unpublished results